\documentclass[prl,twocolumn,superscriptaddress,nopacs,amsmath,amssymb]{revtex4}
\usepackage{graphicx}  
\usepackage{dcolumn}  
\usepackage{bm}  
\usepackage{epstopdf}
\usepackage{times}
\usepackage{mathptmx}
\usepackage{setspace}

\begin{document}

\title{Exchange-driven magnetoresistance in silicon facilitated by electrical spin injection}

\author{Yuichiro~Ando}
\affiliation{Dept. of Electronics, Kyushu University, 744 Motooka, Fukuoka 819-0395, Japan}

\author{Lan~Qing}
\affiliation{Dept. of Physics and Astronomy, University of Rochester, Rochester, New York 14627, USA}

\author{Yang~Song}
\affiliation{Dept. of Electrical and Computer Engineering, University of Rochester, Rochester, New York 14627, USA}

\author{Shinya~Yamada}
\affiliation{Dept. of Electronics, Kyushu University, 744 Motooka, Fukuoka 819-0395, Japan}
	
\author{Kenji~Kasahara}
\affiliation{Dept. of Electronics, Kyushu University, 744 Motooka, Fukuoka 819-0395, Japan}

\author{Kentarou~Sawano}
\affiliation{Advanced Research Laboratories, Tokyo City University, 8-15-1 Todoroki, Tokyo 158-0082, Japan}

\author{Masanobu~Miyao}
\affiliation{Dept. of Electronics, Kyushu University, 744 Motooka, Fukuoka 819-0395, Japan}

\author{Hanan~Dery}
\altaffiliation{hanan.dery@rochester.edu}
\affiliation{Dept. of Physics and Astronomy, University of Rochester, Rochester, New York 14627, USA}
\affiliation{Dept. of Electrical and Computer Engineering, University of Rochester, Rochester, New York 14627, USA}

\author{Kohei~Hamaya}
\altaffiliation{hamaya@ed.kyushu-u.ac.jp}
\affiliation{Dept. of Electronics, Kyushu University, 744 Motooka, Fukuoka 819-0395, Japan}

\begin{abstract} 
We use electrical spin injection to probe exchange interactions in phosphorus doped silicon (Si:P). The detection is enabled by a magnetoresistance effect that demonstrates the efficiency of exchange in imprinting spin information from the magnetic lead onto the localized moments in the Si:P region. A unique Lorentzian-shaped signal existing only at low temperatures ($\lesssim 25 K$) is observed experimentally and analyzed theoretically in electrical Hanle effect measurement. It stems from spin-dependent scattering of electrons by neutral impurities in Si:P. The shape of this signal is not directly related to spin relaxation but to exchange interaction between spin-polarized electrons that are localized on adjacent impurities.
\end{abstract}

\maketitle

The exchange interaction between free and localized electrons is the driving force for the magnetic phases of rare-earth compounds \cite{Shishido_Science10}, diluted magnetic semiconductors \cite{Dietl_Science00}, metallic spin glasses \cite{Mezard_B87}, and oxide interfaces \cite{Brinkman_NatMater07}. In quantum computing architectures, this interaction is a viable candidate for state preparation and setting up entanglement \cite{Bayat_PRL12}. In semiconductor spintronics, however, it is not manifested by straightforward electrical detection due to the lack of intrinsic magnetic interactions in devices made of nonmagnetic semiconductors. Whereas such devices enable long spin transport between the injection and detection terminals on accounts of the relatively weak spin-orbit coupling \cite{Crooker_Science05,Appelbaum_Nature07,Li_PRL13}, they still lack viable means to manipulate the spin transport.

In this Letter, we report a robust exchange effect in a silicon-based spintronic device wherein the mobility of free electrons and the imprinted spin information onto localized moments are dictated by this interaction. The detection is made via a magnetoresistance effect facilitated by electrical spin injection from a ferromagnetic lead and by doping the silicon on the verge of its critical insulator-to-metal transition. By mapping the dependence of the exchange-driven voltage signal on temperature, electric and magnetic fields, we are able to distinguish it from electrical signatures of bare spin accumulation \cite{Crooker_Science05,Appelbaum_Nature07,Li_PRL13,Jonker_Nature07,Suzuki_APE11,Sasaki_APL11,Ando_APL11b} or impurity-assisted tunneling magnetoresistance \cite{Song_arXiv14,Tran_PRL09}. Meanwhile, we show that electrical spin injection to a region populated by $\sim$$\,$10$^{11}$~cm$^{-3}$ free electrons in steady-state can measurably polarize a population of  $>$10$^{17}$~cm$^{-3}$ localized electrons. The exchange mechanism offers new functionalities for spin-based silicon devices such as the control of electron mobility and readout of spin information.

\begin{figure}
\includegraphics[width=8.5cm]{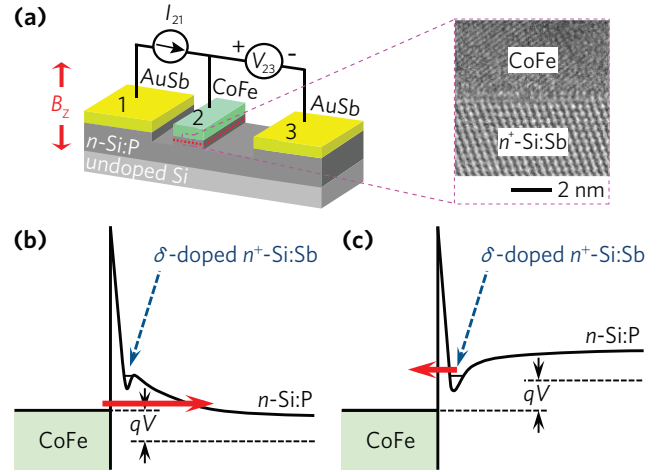}
\caption{(Color online). Device geometry and energy band profile of the CoFe/Si junction. (a) Schematic diagram of the three-terminal device. The enlarged picture is a cross-sectional image of the magnetic heterointerface taken by transmission electron microscopy.  (b)-(c) Cartoons of tunneling in spin injection and extraction respectively.  The doping-induced potential well is not involved in injection unless the temperature increases. In extraction, the current is governed by escape from the metastable state of the well. }
\label{fig:scheme}
\end{figure}

Figure~\ref{fig:scheme}(a) shows a scheme of the employed three-terminal device and a cross-sectional image of its CoFe/Si heterointerface. The substrate is a commercial undoped float-zone Si(111) followed by 200~nm phosphorus-doped silicon (Si:P) where $N_\text{P}$$\,$$\sim$$\,$6$\times$10$^{17}$~cm$^{-3}$ \cite{Ando_PRB12}. The area of the Schottky junction between the magnetic lead (CoFe; contact 2) and the Si part is 200$\,$$\times$$\,$6~$\mu$m$^2$.  The 5$\,$-$\,$7~nm below the magnetic contact is a heavily antimony-doped silicon (Si:Sb) where $N_\text{Sb}$$\,$$\sim$$\,$2$\times$10$^{19}$~cm$^{-3}$ \cite{Sawano_APL10,Maeda_APL10}.  Figures~\ref{fig:scheme}(b) and (c) show cartoons of the resulting energy band profile in the CoFe/Si junction under conditions of spin injection and extraction, respectively (electrons flow from and into the CoFe lead). To perform the measurements, two nonmagnetic ohmic terminals (AuSb; contacts 1 and 3) are fabricated $\sim$50/70~$\mu$m to the left/right of the magnetic lead. The detected voltage signal is the change in $V_{23}$ in response to application of a weak out-of-plane magnetic field while the spin-injection current $I_{21}$ is held constant. Further details on the fabrication and measurement procedures are given in the supplemental material \cite{supple}.

\begin{figure*}
\includegraphics[width=17.5cm]{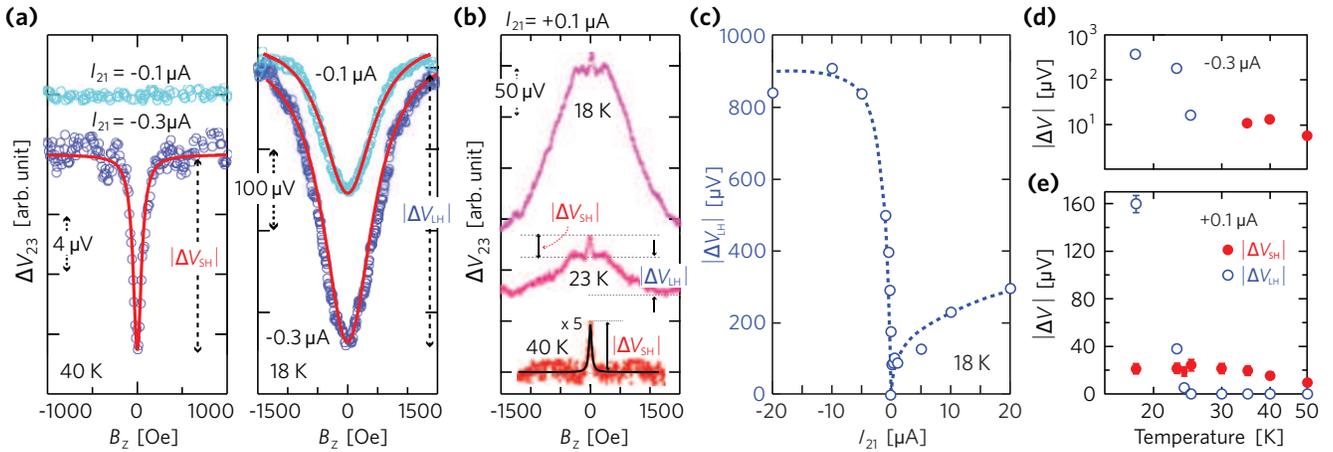}
\caption{(Color online). Main experimental findings. (a) $\Delta$$V_{23}$-$B_\text{Z}$ curves in spin injection at 40 and 18~K.  (b) $\Delta$$V_{23}$-$B_\text{Z}$ curves in spin extraction at 40, 23, and 18~K. The solid curves are Lorentzian-fitted functions. (c) Amplitude of the large signal versus the current at 18~K. (d)-(e) Amplitude of the large (open symbols) and small (solid symbols) signals versus temperature for $I_{21}$$\,$$=$$\,$$-$0.3~$\mu$A and $+$0.1~$\mu$A, respectively.}
\label{fig:exp}
\end{figure*}

We summarize the main experimental findings. Figure~\ref{fig:exp}(a) shows the detected signals at 18~K and 40~K for spin injection. At 18~K and $-0.3$~$\mu$A, the amplitude and halfwidth of the signal are of the order of $\sim$$\,$350~$\mu$V and $\sim$$\,$500~Oe, respectively.  At 40~K, on the other hand, the respective values are $\sim$10~$\mu$V and $\sim$$\,$50~Oe.  As shown in Fig.~\ref{fig:exp}(b) for spin extraction at $+0.1$~$\mu$A, both signals coexist in the low-field region below $\sim$$\,$23~K, while only the smaller signal survives at higher temperatures.  To augment these findings, Fig.~\ref{fig:exp}(c) shows the amplitude dependence of the larger signal on the current at 18~K. Figures~\ref{fig:exp}(d) and (e) show the amplitude dependence of both signals on temperatures at injection and extraction levels of $I_{21} = -0.3$~$\mu$A  and $+0.1$~$\mu$A, respectively. All of the voltage signals were observed for modulation with out-of-plane magnetic fields. We did not observe modulation with in-plane field which is a typical attribute of impurities in the tunnel barrier \cite{Song_arXiv14,Tran_PRL09}, or of stray fields due to interface roughness \cite{Dash_PRB11}. The lack of in-plane field modulation in our direct-contact device is reasoned by the high-quality heterointerface [Fig.~\ref{fig:exp}(a)]. 

The distinctive amplitudes and halfwidths of the large and small signals indicate that their underlying physics is different. To analyze these observations we first check if the measured signals reflect simple spin accumulation in the Si region. In this case, the Lorentzian-shaped  signal  is reminiscent of the Hanle effect in optical spin injection \cite{Zutic_RMP04}, and can be tested by the relation $(\gamma_e \tau_s)^{-1}  \sim  \delta B$ between the spin relaxation time ($\tau_s$) and the measured halfwidth ($\delta B$), where $\gamma_e$$\,$$\approx$$\,$1.7$\times$10$^{7}$~s$^{-1}$$\cdot$Oe$^{-1}$ is the electron gyromagnetic ratio in Si. The spin relaxation times of free electrons in the conduction band and of localized electrons on donor sites are well known from electron paramagnetic resonance experiments (EPR) \cite{Pifer_PRB75,Castner_PR63,Zarifis_PRB98}. In the Si:P region of our device [$N_\text{P}$$\,$$\sim$$\,$6$\times$10$^{17}$~cm$^{-3}$; see Fig.~\ref{fig:scheme}(a)], the spin relaxation of free electrons is in the ballpark of  $\sim$$\,$100~ns below 100~K and showing a weak  T-dependence \cite{Pifer_PRB75}. The spin relaxation time of localized electrons drops from 800~ns at 18~K to 14~ns at 40~K \cite{Castner_PR63}. We realize, therefore, that spin relaxation in Si:P can support $\delta B$$\,$$<$$\,$5~Oe which is far narrower than the halfwidths of both detected signals.  In contrast to the long spin relaxation in Si:P, $\sim$1~ns is a viable spin lifetime in the interface region of our device which comprises of heavily Sb-doped silicon [$N_\text{Sb}$$\,$$\sim$$\,$2$\times$10$^{19}$~cm$^{-3}$; see Fig.~\ref{fig:scheme}(a)]. This timescale is supported by EPR findings \cite{Zarifis_PRB98,Pifer_PRB75}, and by our previous findings in devices with similar Si:Sb interface but with different doping concentrations in the Si:P region \cite{Ando_APL11a,Ando_APL11b}.  

To further support that the smaller Hanle signal ($\Delta V_\text{SH}$ with $\delta B$$\,$$\sim$$\,$50~Oe) stems from spin accumulation in the interface region, we focus on the energy band profile of the CoFe/Si junction. As illustrated by Fig.~\ref{fig:scheme}(b), the Schottky barrier extends into a small portion of the Si:P region in injection conditions \cite{footnote}. Thus, tunneling from the CoFe lead is carried directly to the Si:P region leaving the potential well empty at low temperatures, where the well is formed by the doping inhomogeneity \cite{Dery_PRL07,Song_PRB10}. Injection via the well is enabled when the temperature increases, and this effect is more meaningful when the depletion region extends further into the Si:P region. This qualitative picture supports the fact that we observe the small signal only above certain temperature and current levels [Figs.~\ref{fig:exp}(a) and (d)].  In extraction, illustrated by Fig.~\ref{fig:scheme}(c), the tunneling picture is different since the current is composed almost entirely by escape of electrons from the potential well into the magnetic lead \cite{Dery_PRL07,Song_PRB10}. This tunneling is temperature-independent, and therefore explains the double Hanle-like feature in extraction below 25~K. That is, the small signal is from spin accumulation in the Si:Sb interface region while the large signal is from the Si:P region as we explain below.

\begin{figure}
\includegraphics[width=8.5cm]{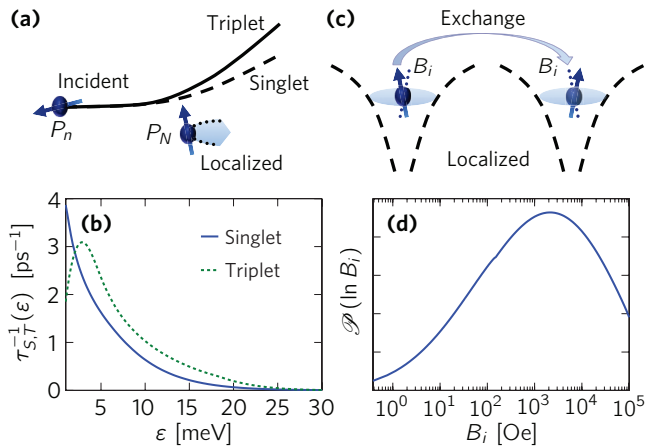}
\caption{(Color online). Exchange interactions in the Si:P region. (a) Trajectory of a free electron due to elastic scattering off a neutral impurity. The scattering amplitude depends on the spin configuration of the free and localized electrons. (b) The resulting scattering rates of free electrons versus their energy for $N_\text{P}$$\,$$=$$\,$6$\times$10$^{17}$~cm$^{-3}$, calculated by the phase shift method \cite{Kwong_PRB98}. (c) Nearest-neighbor exchange coupling. A nonzero spin polarization of localized electrons builds up an internal magnetic field. (d) Distribution of internal fields for $4\pi N_\text{P}a_B^3/3 = 0.02$ \cite{Andres_PRB81}. We use a logarithmic scale due to the large range of exchange couplings, $\mathcal{P}(\ln{B})$$\,$$=$$\,$$B$$\mathcal{P}(B)$.} \label{fig:theory1}
\end{figure}

Turning to the central result of this work, we analyze the larger Hanle signal ($\Delta V_\text{LH}$ with $\delta B$$\,$$\sim$$\,$500~Oe) showing that it stems from exchange interactions in the Si:P region. These interactions are manifested in three ways. The first is via exchanging the spins of injected free electrons and  localized electrons \cite{Honig_PRL66,Mahan_PRB08}. This process imprints the spin information of the magnetic lead onto the localized moments, enabled by employing a doping concentration that keeps the silicon in its insulating phase at low temperatures. As a result, most of the electrons in the Si:P region freeze out on donors while the steady-state minute portion of free electrons is injected from the magnetic lead. The second exchange manifestation is via scattering of the injected free electrons off the neutral impurities that host the localized electrons. As illustrated by  Fig.~\ref{fig:theory1}(a) and quantified in Fig.~\ref{fig:theory1}(b), the scattering amplitude depends on whether the free and localized electrons share a triplet or singlet spin state \cite{Honig_PRL66,Ghosh_PRB92,Kwong_PRB98}. In the absence of spin polarization, the probability to have triplet rather than singlet scattering is three times larger on accounts of degeneracy considerations. Electrical spin injection, however, increases this ratio to $(3+P_nP_N):(1-P_nP_N)$, where $P_n$ and $P_N$ are the spin polarization of free and localized electrons, respectively. Application of an out-of-plane magnetic field induces spin precession, and therefore decreases the ratio toward the unpolarized value (3:1). When the transport is governed by scattering off neutral impurities, the net result is that a weak external magnetic field can effectively modulate the mobility of electrons where the effect is commensurate with $P_nP_N$. The signal amplitude, $\Delta V_\text{LH}$, is a measure of the electric-field change in the Si:P region under the magnetic  lead. Given that the current between contacts 1 and 2 is fixed,  the drift velocity of electrons remains  unaffected if the change in amplitude of the electric field is inversely proportional to that of the mobility. The third manifestation is the exchange coupling between localized electrons on nearby neutral donors. This exchange gives rise to a net internal magnetic field due to the spin polarization of electrons ($P_N \neq 0$). The internal field points along the magnetization axis of the magnetic lead [see Fig.~\ref{fig:theory1}(c)], and as a result, depolarization by spin precession becomes viable only when the out-of-plane external field is comparable or larger than the in-plane internal field. This effect sets the width of the large signal via the magnetic-field dependence of $P_N$. Figure~\ref{fig:theory1}(d) shows the nearest-neighbor exchange couplings distribution in our Si:P region \cite{Andres_PRB81}. It is governed by the localization length (Bohr radius of Si) and the statistics of the inter-donor distance \cite{supple}.

\begin{figure}[t]
\includegraphics[width=7.5cm]{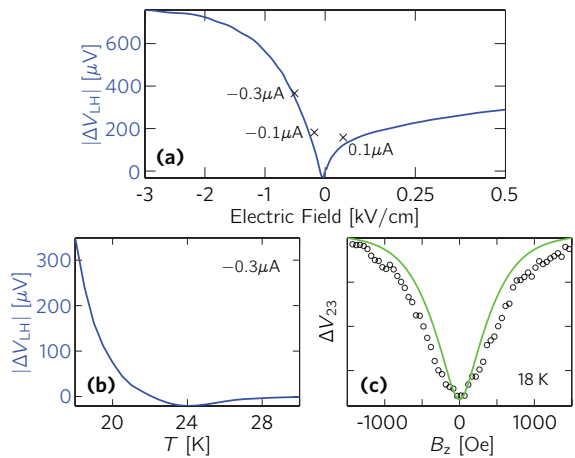}
\caption{ (Color online). Theoretical analysis of the large $\Delta$$V_{23}$ signal for $P_J$$\,$$=$$\,$0.8. (a), Signal amplitude versus the $E$ field at 18~K. Cross marks denote the experimental findings of Fig.~\ref{fig:exp}. (b), Signal amplitude versus temperature for $I_{21} \sim -0.3\mu$A.  (c),  Out-of-plane $B$ field dependence. The solid line and black circles are the theoretical and experimental findings in $I_{21}=-0.3~$$\mu$A, respectively.}
\label{fig:theory2}
\end{figure}

Considering these three exchange effects together we can reproduce all the experimental trends of the large Lorentzian-shaped signal. Figure~\ref{fig:theory2}(a) shows the calculated amplitude as a function of the $E$ field at 18 K. It follows from
\begin{equation}
\Delta V_{\text{LH}} = \frac{ (\bar{\tau}_T  -\bar{\tau}_S) E \ell}{3\bar{\tau}_S+\bar{\tau}_T+\tau_0}  P_{n,0}P_{N,0}\,\,,\label{eq:DV}
\end{equation}
where $\bar{\tau}_{T}$ ($\bar{\tau}_{S}$) are the average neutral-impurity scattering time for the triplet (singlet) spin configuration, $\tau_0 \equiv 4\bar{\tau}_T\bar{\tau}_S/\bar{\tau}_m$, and $\bar{\tau}_m$ is the average momentum relaxation time due to all other scattering mechanisms. The average values of the scattering times are extracted from Monte Carlo simulations that consider the $E$-field dependent distribution of hot electrons in the Si:P region \cite{supple,Li_PRL12}. Other parameters in Eq.~(\ref{eq:DV}) are the spin polarizations of free and localized electrons, evaluated at zero external magnetic field, and $\ell$ which is the effective transport length under contact 2. This length scale is set by the nominal thickness of the Si:P layer ($\ell$$\,$$=$$\,$200 nm) due to the fact that at low temperatures, injected free electrons cross the Si:P region to the substrate where the flow between contacts 1 and 2 is far less resistive (free of impurity scattering) \cite{Appelbaum_Nature07}. The fast transit time to reach the substrate (e.g., 10~ps at 18~K when $I$$\sim$$-$3~$\mu$A), means that the spin polarization of the free electrons matches that of the injected current, $P_n \sim \pm P_J$, where the $+$/$-$ sign denotes injection/extraction. The theoretical  results in Fig.~\ref{fig:theory2}(a) are displayed as a function of the $E$-field since the conversion to current levels at large fields is complicated by violation of charge neutrality in freeze out conditions. Nonetheless, we readily recognize the agreement in nonlinear behavior and amplitude scales between Fig.~\ref{fig:theory2}(a) and the experimental findings in Fig.~\ref{fig:exp}(c). The different trends in injection ($E<0$) and extraction ($E>0$) are caused by the asymmetric role of the interface and by the fact that in strong extraction, the $E$ field hinders the spin diffusion away from the junction. 

Next, we note that the calculated temperature dependence of $\Delta V_{\text{LH}}$, shown in Fig.~\ref{fig:theory2}(b) for $I_{21} \sim -0.3$~$\mu$A, matches the experimental findings in Fig.~\ref{fig:exp}(d). This dependence is governed by the exponential increase of the conductivity in the Si:P region which commensurate with the density of free electrons ($n_e$). We can understand it by noting that $n_e \approx \sqrt{N_\text{P}N_C}\exp{(-E_0/2k_BT)}$, where $E_0$$\,$$\sim$$\,$45~meV is the donor ionization energy and $N_C\simeq2.6 \times 10^{15} \times  T^{3/2}$~cm$^{-3}$ is the effective density of states in the conduction band.  In our device,  $n_e$ increases from $\sim$$\,$2$\times$10$^{11}$~cm$^{-3}$  at 18~K to $\sim$$\,$2$\times$10$^{15}$~cm$^{-3}$ at 40~K, whereas the density of localized electrons is nearly $N_\text{P}$ across this temperature range. The exponential increase of $n_e$ is accompanied by an inverse decrease of the electric field, giving rise to a strong suppression of $\Delta V_{\text{LH}}$ [$E \propto I_{21}/n_e$ in Eq.~(\ref{eq:DV})]. Finally, Fig.~\ref{fig:theory2}(c) shows the normalized value of $\Delta V_{23}$ against $B_z$ along with the experimental findings in $-$0.3~$\mu$A.  The statistical nearest-neighbor exchange model captures the measured $B$ field dependence. This result clearly shows that the Lorentzian shape, often detected in electrical spin injection experiments, is not necessarily governed by spin relaxation. Further theoretical and calculation details are provided in the supplemental material \cite{supple}.

The findings of this work demonstrate the power of electrical spin injection in probing subtle spin interactions in nonmagnetic materials. Ascribing the exchange mechanism to the large Lorentzian-shaped signal in our device is facilitated by two important factors. First, we employ a direct Schottky contact between CoFe and Si, thereby eliminating spurious magnetoresistance effects due to impurities embedded in oxide tunnel barriers \cite{Song_arXiv14,Tran_PRL09}. Second, choosing the doping concentration to be on the verge of the insulator-to-metal critical transition in Si ($N_\text{P}=6\times10^{17}$~cm$^{-3} < 2\times10^{18}$~cm$^{-3}$) ensures that at low temperatures most of the electrons are still localized on donor sites while their exchange coupling is maximal. Taking these features into account together with the highly nonlinear dependence of the larger voltage signal on the temperature and current, we can clearly set apart our findings from those of three-terminal devices that employ oxide barriers and a highly-degenerate semiconductor (i.e., metallic) \cite{Dash_Nature09,Li_NatureComm11, Gray_APL11,Jain_PRL12,Jeon_PRB13,Pu_APL13}. In such devices, the magnetoresistance effect can stem from the spin polarization of electrons in oxide defects for which the temperature dependence is measurably weaker \cite{Uemura_APL12,Txoperena_APL13}. In closing, we remark that the development of semiconductor spin-based logic devices remains a great challenge \cite{Zutic_RMP04,Dery_Nature07}. This work, however, shows that exchange offers new possibilities to modulate the charge mobility, to effectively imprint the spin polarization onto localized moments in semiconductors, and to enable readout of the spin information. Integrating these ingredients in spin-based devices made of silicon,  the ubiquitous material in the microelectronic industry, is a viable route to development of semiconductor spintronics.

The work in Japan is partly supported by PRESTO-JST and KAKENHI from JSPS (Contract No. 25246020). The work in USA is supported by NRI-NSF, NSF, and DTRA (Contracts No. DMR-1124601, ECCS-1231570, and HDTRA1-13-1-0013). Y.A. K.K. and S.Y. acknowledge JSPS Research Fellowships for Young Scientists.

\end{document}